\documentclass[reprint,prl,twocolumn,superscriptaddress,showpacs,showkeys,floatfix,preprintnumbers]{revtex4-2}
\usepackage[english]{babel}
\usepackage[utf8]{inputenc}
\usepackage[T1]{fontenc}
\usepackage{amsmath,amssymb,braket,slashed,mathrsfs}
\usepackage{pifont}
\usepackage{hyperref}
\usepackage{color,hyperref}
\usepackage{cleveref}
\usepackage{amsmath}
\usepackage{graphicx}
\usepackage{booktabs}
\usepackage{multirow}
\usepackage{subfigure}
\usepackage{float}

\newcommand{\beq}{\begin{eqnarray}}
\newcommand{\eeq}{\end{eqnarray}}
\newcommand{\be}{\begin{equation}}
\newcommand{\ee}{\end{equation}}
\newcommand{\ba}{\begin{eqnarray}}
\newcommand{\ea}{\end{eqnarray}}
\newcommand{\nn}{\nonumber}
\newcommand{\bit}{\begin{itemize}}
\newcommand{\eit}{\end{itemize}}
\newcommand{\rw}{\rightarrow}

\newcommand{\zzuphy}{School of Physics and Microelectronics, Zhengzhou University, Zhengzhou, Henan 450001, China}
\newcommand{\innovation}{Collaborative Innovation Center of Quantum Matter, Beijing 100871, China}
\newcommand{\chep}{Center for High Energy Physics, Peking University, Beijing 100871, China}
\newcommand{\pkuphy}{School of Physics, Peking University, Beijing 100871,
China}
\newcommand{\KeyLab}{State Key Laboratory of Nuclear Physics and Technology,
Peking University, Beijing 100871, China}

\begin{document}

\title{First-principle calculation of the $\eta_c \rw 2\gamma$ decay width from lattice QCD}

\author{Yu Meng}
\email[Email: ]{yu_meng@zzu.edu.cn}
\affiliation{\zzuphy}
\author{Xu Feng}
\email[Email: ]{xu.feng@pku.edu.cn}
\affiliation{\pkuphy}\affiliation{\chep}\affiliation{\innovation}
\affiliation{\KeyLab}
\author{Chuan Liu}\affiliation{\pkuphy}\affiliation{\chep}\affiliation{\innovation}
\author{Teng Wang}\affiliation{\pkuphy}
\author{Zuoheng Zou}\affiliation{\pkuphy}

\date{\today}

\begin{abstract}
We perform a lattice QCD calculation of the $\eta_c\to2\gamma$ decay width
using a model-independent method that requires no momentum extrapolation
of the off-shell form factors.
This method also provides a straightforward and simple way to examine the finite-volume effects.
The calculation is accomplished using $N_f=2$ twisted mass fermion ensembles. The statistically significant excited-state effects are observed and eliminated using a multi-state fit.
The impact of fine-tuning the charm quark mass is also examined
and confirmed to be well-controlled.
Finally, using three lattice spacings for the continuum extrapolation, we obtain the decay width
$\Gamma_{\eta_c\gamma\gamma}=6.67(16)_{\mathrm{stat}}(6)_{\mathrm{syst}}$ keV,
which differs significantly from the Particle Data Group's reported value of $\Gamma_{\eta_c\gamma\gamma}=5.4(4)$ keV (2.9~$\sigma$ tension).
We provide insight into the comparison between our findings, previous theoretical predictions, and experimental measurements.


\end{abstract}

\maketitle

\section{Introduction}
As a multi-scale system that can probe various regimes of quantum chromodynamics (QCD), heavy quarkonium presents an ideal laboratory
for testing the interplay between perturbative and nonperturbative QCD~\cite{Brambilla:2010cs}.
In quarkonium physics, the two-photon decay widths of quarkonium play an important role in connecting QCD from perturbative to nonperturbative regime.
These quantities are traditionally expressed as the product of the short-distance quark-antiquark annihilation decay rates and the
squared bound-state wave function at the origin~\cite{Bodwin:1994jh}. Phenomenologically, the latter provides an essential, universal input
for calculating the decay and production cross sections for the quarkonium states~\cite{Eichten:1995ch}.
Therefore, quarkonium physics relies heavily on the accurate determination of these decay widths.

In this study, we focus on examining the two-photon decay of the lowest charmonium state, $\eta_c\to2\gamma$, a topic that has attracted extensive
attention from both experimental~\cite{PLUTO:1985fgr,Turin:1987qkd,TPCTwoGamma:1987tnp,CLEO:1989cel,L3:1993ahi,ARGUS:1994glp,E760:1995rep,AMY:1998ghf,L3:1999byj,CLEO:2000qcl,FermilabE835:2003ula,DELPHI:2003kmy,CLEO:2003gwz,Belle:2005fji,BaBar:2005pcw,Belle:2012uhr,BESIII:2012lxx,CLEO:2008qfy}
and theoretical studies~\cite{Godfrey:1985xj,Bodwin:1994jh,Dudek:2006ut,Feng:2015uha,CLQCD:2016ugl,Chen:2016bpj,Feng:2017hlu,Li:2019ncs,Liu:2020qfz,Zhang:2021xvl}.
On the experimental side, the low statistics for direct measurements make the accurate determination of the two-photon decay extremely difficult.
Despite decades of effort, direct measurements of decay width still have uncertainties ranging from 20\% to 100\%.
The decay width $\Gamma_{\eta_c\gamma\gamma}=5.4(4)$ keV favored by the Particle Data Group (PDG) is compiled using a combined fit with other decay channels, resulting in a branching ratio of $\operatorname{Br}(\eta_c\rw 2\gamma)=(1.68\pm 0.12)\times 10^{-4}$,
denoted here as the ``PDG-fit'' value.
However, if one examines the PDG list in detail, there is another value of
$\operatorname{Br}(\eta_c\rw 2\gamma)=2.2^{+0.9}_{-0.6}\times 10^{-4}$ compiled based on the
average of BESIII~\cite{BESIII:2012lxx} and CLEO~\cite{CLEO:2008qfy} measurements,
with $\operatorname{Br}(\eta_c\rw2\gamma)=(2.7\pm0.8\pm0.6)\times10^{-4}$
and $\operatorname{Br}(\eta_c\rw2\gamma)=(0.7^{+1.6}_{-0.7}\pm0.2)\times10^{-4}$, respectively.
These results are rather far apart from each other, but still consistent due to large errors.
Such large errors propagate into the PDG-aver value and result in a 28\% uncertainty.
We denote this value as the ``PDG-aver'' value.
Notably, PDG-fit value has a 7 times smaller uncertainty and a 24\% lower central value
compared to the PDG-aver one. As for the PDG-fit value, other decay channels are also taken into account.
The constraints from different $\eta_c$ decays result in a much smaller error
and a downward shift of the central value.
The significantly suppressed uncertainty in the PDG-fit value indicates that
the direct experimental measurements of the $\eta_c\rw2\gamma$ decay have little impact here.

On the theoretical side, a recent calculation based on Dyson-Schwinger equation~\cite{Chen:2016bpj}
suggests a two-photon decay width
$\Gamma_{\eta_c\gamma\gamma}=6.32 \textrm{\textendash} 6.39$ keV (with a branching ratio of $\operatorname{Br}(\eta_c\rw 2\gamma)=(1.98 \textrm{\textendash} 2.00)\times10^{-4}$),
consistent with the PDG-aver value but much larger than the PDG-fit one.
Meanwhile, a study from nonrelativistic QCD, including the next-to-next-to-leading-order perturbative corrections~\cite{Feng:2017hlu}, gives the branching ratio
$\operatorname{Br}(\eta_c\rw 2\gamma)=(3.1\textrm{\textendash}3.3)\times10^{-4}$, which is larger than other theoretical predictions and the experimental measurements
To clarify the discrepancies, a first-principle calculation of $\eta_c\to2\gamma$ decay width from lattice QCD is crucial.

Lattice QCD calculations of $\eta_c \rw 2\gamma$ decay have been conducted in the past,
but the systematic effects in these computations are still not entirely under control.
Apart from the earlier quenched studies~\cite{Dudek:2006ut},
later unquenched lattice studies have utilized only one or two
lattice ensembles~\cite{CLQCD:2016ugl,Liu:2020qfz}. In particular,
substantial lattice spacing errors have been observed in Ref.~\cite{Liu:2020qfz}. Therefore,
it is crucial to use at least three lattice ensembles to allow for a controlled continuum extrapolation.
This work aims to systematically improve upon previous lattice studies of this radiative decay.
Several improvements are made to obtain a more accurate result: (1) We adopt a novel method to extract the on-shell form factor directly,
which avoids conventional model-dependent errors caused by the momentum extrapolation
of the off-shell form factors.
(2) We perform a spatial volume integral to obtain the form factor, with a
truncation range introduced to monitor the finite-volume effects.
(3) We remove the excited-state contaminations,
which are found to be quite significant in this study.
(4) We confirm that the systematic effects due to fine-tuning the valence charm quark mass
are smaller than the statistical errors.
(5) Lastly, lattice computations are done using three ensembles with
three different lattice spacing $a$.
We finally perform a continuum extrapolation using three ensembles with different lattice spacings and find
that the lattice results are well described by a form linear in $a^2$,
which is suggested by the automatic $O(a)$ improvement of the ensembles.
These efforts finally allow us to obtain the decay width with a precision of about 2.6\%.

\section{Methodology}
We start the discussion of $H\to\gamma\gamma$ in an infinite-volume continuum Euclidean space,
where $H$ indicates a hadron with mass $m_H$.
The relevant hadronic matrix element $\mathcal{F}_{\mu\nu}(p)$ for the two-photon decay process is,
\be\label{eq:F_define}
\mathcal{F}_{\mu\nu}(p)=\int \mathrm{d}t\,\mathrm{e}^{m_Ht/2}\int \mathrm{d}^3 \vec{x}\,
\mathrm{e}^{-i\vec{p}\cdot \vec{x}}\mathcal{H}_{\mu\nu}(t,\vec{x}),
\ee
where the hadronic function $\mathcal{H}_{\mu\nu}(t,\vec{x})$ is defined as
\be\label{eq:hadron_3pt}
\mathcal{H}_{\mu\nu}(t,\vec{x})=\langle 0|\textrm{T}[J_{\mu}^{em}(x)J_{\nu}^{em}(0)]|H(k)\rangle ,
\ee
with the initial state $|H(k)\rangle$ carrying the four-momentum $k=(im_H,\vec{0})$. The four-momentum assigned to the electromagnetic current $J_\mu^{em}=\sum_qe_q\,\bar{q}\gamma_\mu q$ ($e_q=2/3,-1/3,-1/3,2/3$ for $q=u,d,s,c$)
takes the form $p=(im_H/2,\vec{p})$ with $|\vec{p}|=m_H/2$, so that it satisfies
the on-shell condition for the photon.

We then assume that the hadron in the initial state is a pseudo-scalar particle. According to its negative parity, the hadronic tensor $\mathcal{F}_{\mu\nu}(p)$ can be parameterized as
\be\label{eq:F_param}
\mathcal{F}_{\mu\nu}(p)=\epsilon_{\mu\nu\alpha \beta}p_{\alpha}k_\beta F_{H\gamma\gamma}.
\ee
By multiplying $\epsilon_{\mu\nu\alpha \beta}p_{\alpha}k_\beta$ to both sides, the form factor at on-shell momentum is
extracted through
\be
F_{H\gamma\gamma}=-\frac{1}{2m_{H}|\vec{\bm{p}}|^2}\int \mathrm{d}^4x\,\mathrm{e}^{-ipx}
\epsilon_{\mu\nu\alpha0}
\frac{\partial \mathcal{H}_{\mu\nu}(x) }{\partial x_{\alpha}}.
\ee
After averaging over the spatial direction for $\vec{p}$, $F_{H\gamma\gamma}$ would be
obtained through
\be\label{eq:F_L}
F_{H\gamma\gamma}=-\frac{1}{2m_{H}}\int \mathrm{d}^4x\, \mathrm{e}^{\frac{m_{H}}{2}t}\,
\frac{j_1(|\vec{\bm{p}}||\vec{\bm{x}}|)}{|\vec{\bm{p}}||\vec{\bm{x}}|}\,
\epsilon_{\mu\nu\alpha0}x_{\alpha}\mathcal{H}_{\mu\nu}(x),
\ee
where $j_n(x)$ are the spherical Bessel functions. The decay width is then given by
\be\label{eq:decay_width}
\Gamma_{H\gamma\gamma}=\frac{\pi}{4}\alpha^2m_{H}^3
F_{H\gamma\gamma}^2.
\ee

In the lattice calculation, we adopt the infinite-volume reconstruction method proposed in Ref.~\cite{Feng:2018qpx}. This method has been
successfully applied to various processes~\cite{Tuo:2019bue,Feng:2019geu,Feng:2020zdc,Christ:2020jlp,Christ:2020hwe,Ma:2021azh,Tuo:2021ewr,Feng:2021zek,Fu:2022fgh,Tuo:2022hft}
to reconstruct the infinite-volume hadronic function
using the finite-volume ones.
In this work, we use it for the lattice calculation of the $\eta_c\rw 2\gamma$ decay.
It is natural to introduce an integral truncation $t_s$ in Eq.~(\ref{eq:F_L}) and write the contribution as
$F_{\eta_c\gamma\gamma}(t_s)$. The parameter $t_s$ is chosen sufficiently large to
guarantee that the time dependence of $\mathcal{H}_{\mu\nu}(t,\vec{x})$ for $t>t_s$ is dominated by the ground state, which
is the $J/\psi$ state when neglecting the disconnected diagrams. Then the residual integral from $t_s$ to $\infty$ can be calculated as
\ba\label{eq:F_INF}
\delta F_{\eta_c\gamma\gamma}(t_s)&=& -\frac{1}{2m_{\eta_c}}\frac{\mathrm{e}^{|\vec{\bm{p}}|t_s}}
{\sqrt{m_{J/\psi}^2+|\vec{\bm{p}}|^2}-|\vec{\bm{p}}|}
\nn\\
&&\times\int \mathrm{d}^3\vec{\bm{x}}\,
\frac{j_1(|\vec{\bm{p}}||\vec{\bm{x}}|)}{|\vec{\bm{p}}||	\vec{\bm{x}}|}
\epsilon_{\mu\nu\alpha0}x_{\alpha}\mathcal{H}_{\mu\nu}(t_s,\vec{\bm{x}}).
\ea
The total contribution of $F_{\eta_c\gamma\gamma}$ is given by
\be
F_{\eta_c\gamma\gamma}=F_{\eta_c\gamma\gamma}(t_s)+\delta F_{\eta_c\gamma\gamma}(t_s),
\ee
where the $t_s$ dependence cancels if the ground state is saturated.
We then use the hadronic function $\mathcal{H}^{L}_{\mu\nu}(t,\vec{x})$ calculated on a finite-volume lattice
to replace the infinite-volume $\mathcal{H}_{\mu\nu}(t,\vec{x})$ for $t\le t_s$.
Such replacement only amounts for exponentially suppressed finite-volume effects as the hadronic function
$\mathcal{H}_{\mu\nu}(t,\vec{x})$ itself is suppressed exponentially when $|\vec{x}|$ becomes large.
One can introduce a spatial integral truncation $R$ and examine at large $R$ whether
the finite-volume effects are well under control or not.

The method used in our calculation is generically different from the conventional approach where
the photon momenta are assigned by discrete Fourier transformation and the off-shell form factors~\cite{Dudek:2006ut} or amplitude squares~\cite{Liu:2020qfz} with nonzero photon virtualities.
In those cases, the physical result can only be obtained after a momentum extrapolation
to the on-shell limit. The situation becomes much easier here as the approach presented above
allows us to extract the on-shell form factor directly.
Therefore, the systematic uncertainties arising from the model-dependent extrapolation are avoided
and the computational cost is also significantly reduced.

\section{Numerical setup}
\begin{table}[!h]
\caption{\label{table:cfgs}%
Parameters of gauge ensembles used in this work. From left to right, we list the ensemble name, the lattice spacing $a$ (with errors taken from Ref.~\cite{Becirevic:2012dc}),
the spatial and temporal lattice size $L$ and $T$, the number of the configurations $N_{\textrm{conf}}$, the light quark mass $a\mu_l$, the corresponding pion mass $m_\pi$ and the range of the time separation between the hadron and the nearest current $\Delta t$,
see the discussion after Eq.~(\ref{eq:th_fit}).
Here, $L$, $T$, and $\Delta t$ are given in lattice units. For all ensembles, $\Delta t$ takes a
a consistent range of 0.7-1.6 fm. }
\vspace{2mm}
\begin{tabular}{ccccccc}
\hline
\textrm{Ensemble} & $a$ (fm) & $L^3\times T$ & $N_{\textrm{conf}}$
& $a\mu_l$ & $m_\pi$ (MeV) & $\Delta t$ \\
%\multicolumn{1}{c}{\textrm{Three}}& \\
\hline
a98 & 0.098(3) & $24^3\times 48$ & $236$ & 0.006 & 365 & 7-16 \\
a85 & 0.085(3) & $24^3\times 48$ & $200$ & 0.004 & 315 & 8-19 \\
a67 & 0.0667(20) & $32^3\times 64$ & $197$ & 0.003 & 300 &10-24 \\
\hline
\end{tabular}
\label{table:cfgs}

\end{table}

The calculation is performed using three $N_f=2$ flavor twisted mass gauge field ensembles generated by
the Extended Twisted Mass Collaboration (ETMC)~\cite{ETM:2009ptp,Becirevic:2012dc} with lattice spacing $a \simeq 0.0667,0.085,0.098$ fm. We call these ensembles a67, a85, and a98, respectively.
More parameters of these ensembles are listed in Table~\ref{table:cfgs}.
The valence charm quark mass is tuned by requiring the lattice result of charmonium masses
to coincide with that of 1) $\eta_c$ or 2) $J/\psi$. These two choices will make a difference on our physical quantities. For simplicity, we add the suffix ``-I'' or ``-II'' to the ensemble name
to specify the case of 1) or 2) mentioned above.
For detailed information on the tuning, we refer to the supplemental material.

In this work, we calculate the three-point correlation function
$C^{(3)}_{\mu\nu}(x,y,t_i)\equiv\langle J^{em}_{\mu}(x)J^{em}_{\nu}(y)\mathcal{O}_{\eta_c}^{\dagger}(t_i)\rangle$
with $t_i=\textrm{min}\{t_x,t_y\}-\Delta t$. The $Z_4$-stochastic wall source propagator is placed at time $t_i$ so that
the $\mathcal{O}_{\eta_c^\dagger}$ operator carries the zero momentum. It is found in our study that the uncertainty is reduced by nearly a factor of 2 by using a stochastic propagator
compared to that using the point source propagator. We also find that the excited-state contamination associated with the $\mathcal{O}_{\eta_c^\dagger}$ operator is significant.
We thus apply the APE~\cite{APE:1987ehd} and Gaussian smearing~\cite{Gusken:1989qx} to the $\eta_c$ field and it efficiently reduces the excited-state effects. Nevertheless,
when the precision reaches 1-3\% in our calculation
the excited-state effects are statistically significant unless $\Delta t \gtrsim 1.6$ fm.
Such systematic effects affect both two-point correlation function $C^{(2)}(t)=\langle \mathcal{O}_{\eta_c}(t) \mathcal{O}_{\eta_c}^{\dagger}(0)\rangle$ and three-point function $C^{(3)}_{\mu\nu}$. Using a two-state fit form, we write the time dependence for $C^{(2)}(t)$ as
\be\label{eq:2pt}
C^{(2)}(t)=V \sum_{i=0,1}\frac{Z_i^2}{2E_i} \left(\textrm{e}^{-E_it}+\textrm{e}^{-E_i(T-t)}\right)
\ee
with $V$ the spatial-volume factor, $E_0=m_{\eta_c}$ the ground-state energy and $E_1$ the energy of the first excited state.
$Z_i=\frac{1}{\sqrt{V}}\langle i|\mathcal{O}_{\eta_c}^\dagger|0\rangle$ ($i=0,1$) are the overlap amplitudes for the ground and the first excited
state. We then use $Z_0$ and $m_{\eta_c}$ as the inputs to determine the hadronic function $\mathcal{H}_{\mu\nu}$
through $\mathcal{H}_{\mu\nu}(t_x-t_y,\vec{x}-\vec{y})=C^{(3)}_{\mu\nu}(x,y,t_i)/[(Z_0/2m_{\eta_c})\textrm{e}^{-m_{\eta_c}(t_y-t_i)}]$.
The excited-state effects carried by $\mathcal{H}_{\mu\nu}$ propagate into the form factor $F_{\eta_c\gamma\gamma}$ and can be parameterized using
a relatively simple two-state form
\be
\label{eq:th_fit}
F_{\eta_c\gamma\gamma}(\Delta t)=F_{\eta_c\gamma\gamma}+\xi\, \textrm{e}^{-(E_1-E_0)\Delta t},
\ee
with two unknown parameters $F_{\eta_c\gamma\gamma}$ and $\xi$.
To fully control the systematic effects, we use various $\Delta t$
from the range of $0.7$ - $1.6$ fm and fit the lattice data to the form~(\ref{eq:th_fit}).
The range of $\Delta t$ in lattice units is also tabulated in the
last column of Table~\ref{table:cfgs}.
To compute the correlation function for the whole set of $\vec{x}-\vec{y}$, we place the point source propagator on one vector current and treat the other one as the sink.
Both the point and the stochastic wall source propagators are placed on all time slices and thus the
average based on time translation invariance can be performed to increase the statistics.

In our calculation, the electromagnetic current is replaced by a local charm quark current as
$J^{em}_\mu(x)=Z_V e_cJ^{(c)}_{\mu}(x)$ with $J^{(c)}_{\mu}$ defined as
$J^{(c)}_{\mu}\equiv \bar{c}\gamma_{\mu}c$. Here $Z_V$ is a renormalization factor,
which converts the local vector current to the conserved one
at the cost of at most of $O(a^2)$.
For the details of the determination of $Z_V$, we refer to the supplemental material.

\section{Lattice results}

\begin{figure}[!h]
\centering
\subfigure{\includegraphics[width=0.47\textwidth]{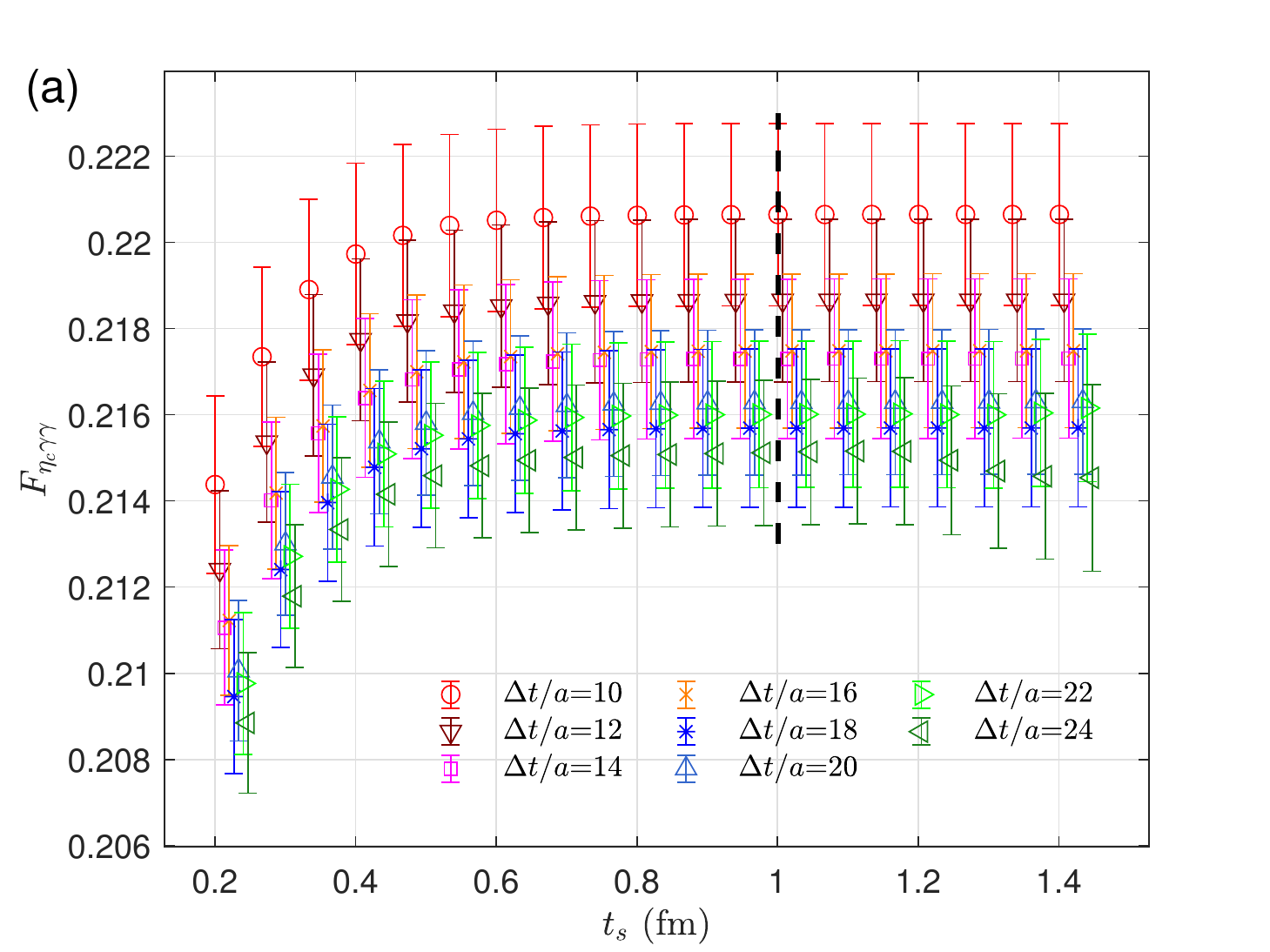}}\hspace{5mm}
\subfigure{\includegraphics[width=0.47\textwidth]{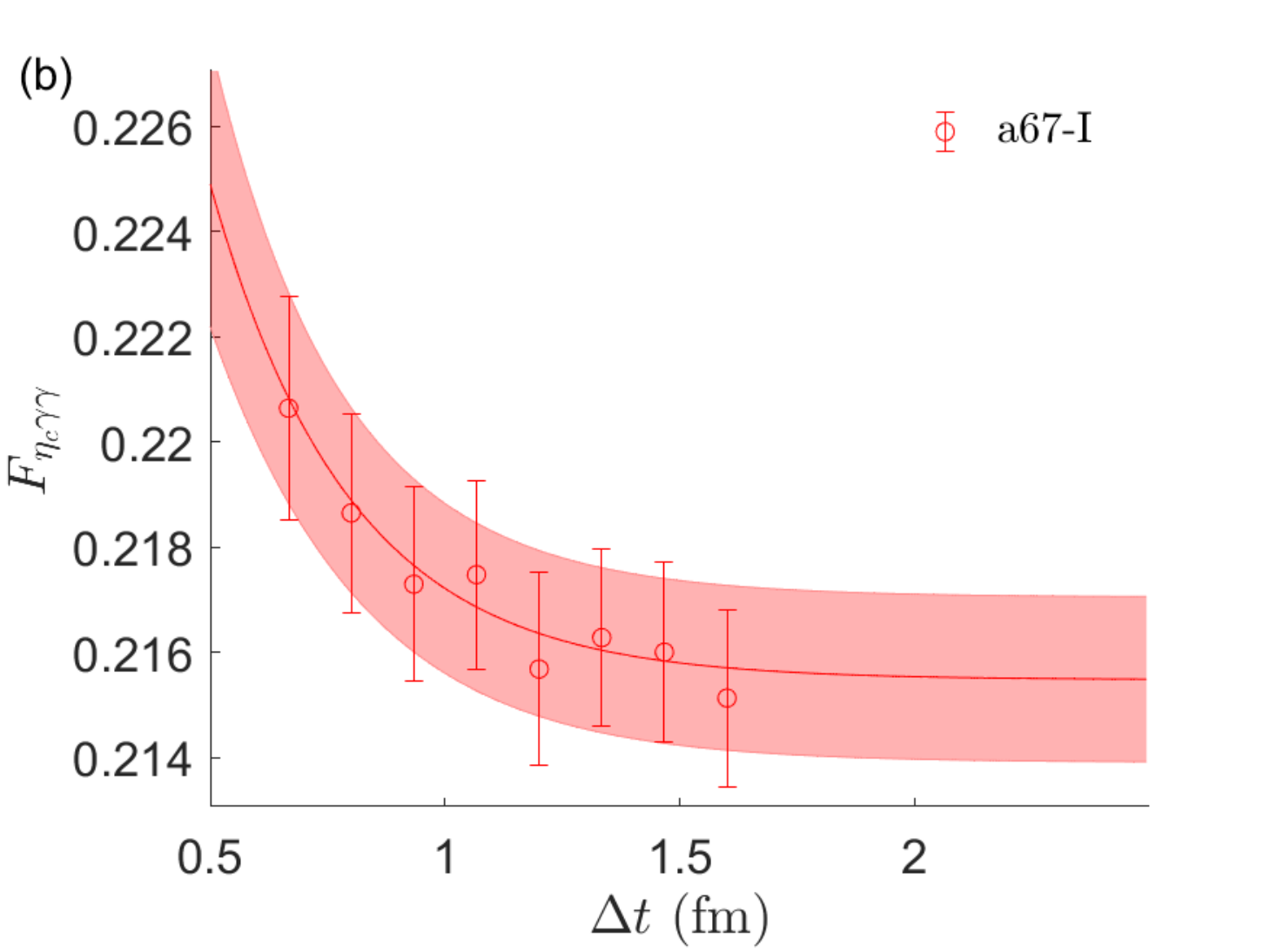}}\hspace{5mm}
\caption{(Color online) The lattice results of $F_{\eta_c\gamma\gamma}$ for ensemble a67-I.
(a) $F_{\eta_c\gamma\gamma}$ is shown as a function of
$t_s$ with various choices of $\Delta t$. The vertical dashed line
denotes a conservative choice of $t_s \simeq 1$ fm,
where the ground-state saturation is realized.
(b) $F_{\eta_c\gamma\gamma}$ with $t_s\simeq 1$ fm
are shown as a function of $\Delta t$ together with a fit to the form~(\ref{eq:th_fit}).}
\label{fig:width_a67}
\end{figure}

The lattice results of $F_{\eta_c\gamma\gamma}$ as a function of the truncation time $t_s$
with different separations $\Delta t$ are shown in the top panel in Fig.~\ref{fig:width_a67}.
Here we take the ensemble with the finest lattice spacing, namely a67 as an example.
The results are shown with the charm quark mass tuned
to the physical point $m_{\eta_c}\simeq m_{\eta_c}^{\mathrm{phys}}$.
The integral in Eq.~(\ref{eq:F_L}) is performed with $\vec{x}$ summed over the whole spatial volume.
We find that for all the separation $\Delta t$ and all ensembles used in this work,
a temporal truncation $t_s\simeq 1$ fm is a conservative choice for the ground-state saturation.
With this choice, the results for $F_{\eta_c\gamma\gamma}$ as a function of $\Delta t$
are shown on the bottom panel in Fig.~\ref{fig:width_a67}.
It shows that $F_{\eta_c\gamma\gamma}$ has an obvious $\Delta t$ dependence,
indicating sizable excited-state effects associated with $\mathcal{O}_{\eta_c}^\dagger$
operator as we have pointed out before.
Using a two-state fit described by Eq.~(\ref{eq:th_fit}) we can extract the ground-state contribution to the form factor at $\Delta t\to\infty$.

\begin{figure}[!h]
\centering
\subfigure{\includegraphics[width=0.47\textwidth]{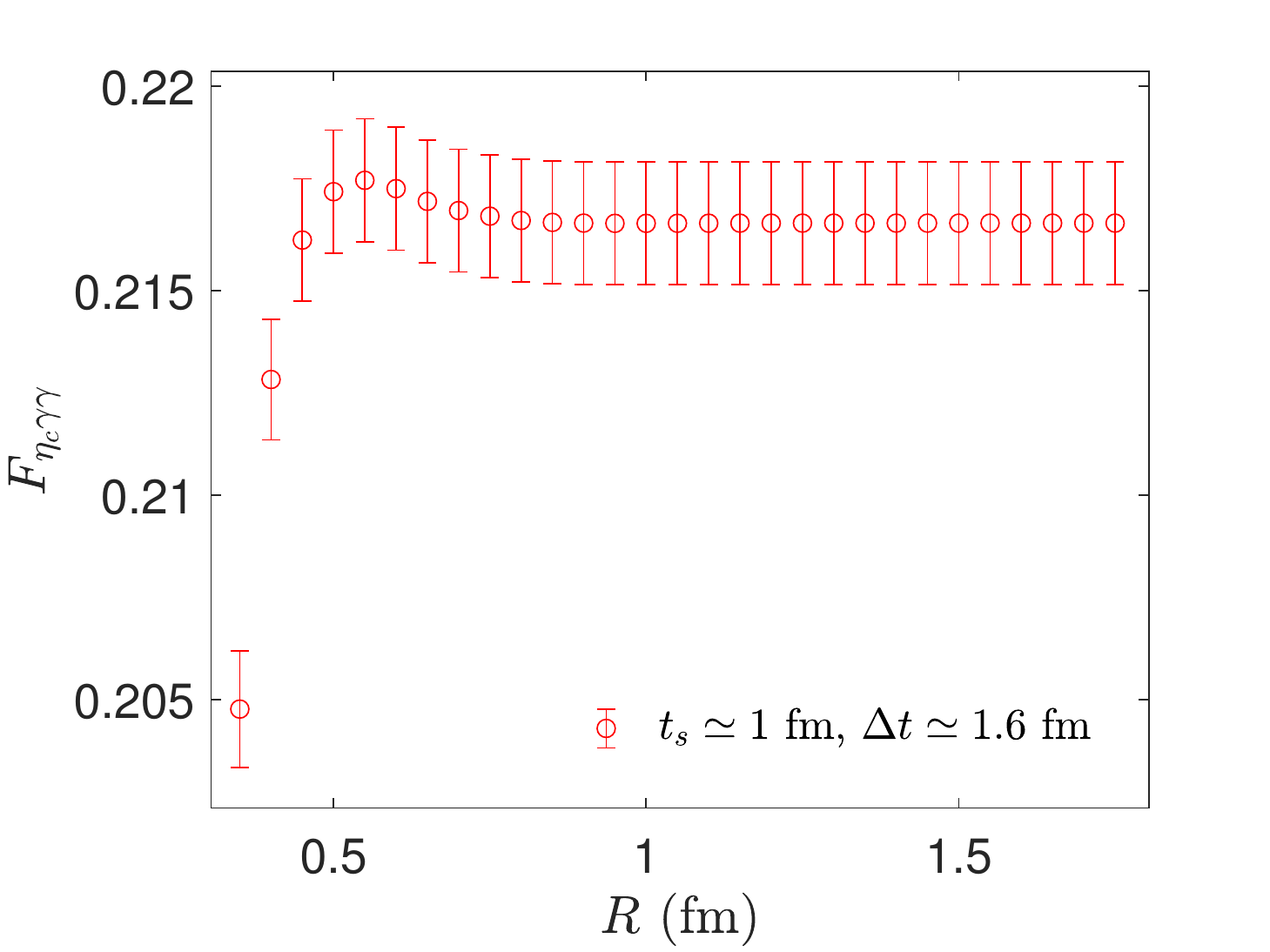}}\hspace{5mm}
\caption{(Color online) For ensemble a67-I, $F_{\eta_c\gamma\gamma}$ with $t_s\simeq 1$ fm and $\Delta t \simeq 1.6$ fm as a function of the spatial range truncation $R$.}
\label{fig:finite}
\end{figure}

To examine the finite-volume effects, we introduce a spatial integral truncation parameter $R$ in both Eqs.~(\ref{eq:F_L}) and (\ref{eq:F_INF}).
As the hadronic function $\mathcal{H}_{\mu\nu}(x)$ is dominated by the $J/\psi$ state at large $|\vec{x}|$, the size of the integrand is exponentially suppressed when $|\vec{x}|$ becomes large.
In Fig.~\ref{fig:finite} the form factor $F_{\eta_c\gamma\gamma}$ is shown as
a function of $R$. For $R\gtrsim 0.8$ fm, there exists a plateau,
indicating that the hadronic function $\mathcal{H}_{\mu\nu}(x)$
at $|\vec{x}|\gtrsim 0.8$ fm has negligible contribution to $F_{\eta_c\gamma\gamma}$.
For each of the three ensembles used in this work,
the lattice size satisfies $L>2$ fm which is sufficiently large to accommodate the hadron.
We thus conclude that finite-volume effects are well under control in our calculation.
Here the parameter $R$ is simply introduced for the examination of
the size of the finite-volume effects.
The lattice results reported throughout the paper are obtained
based on the whole spatial-volume summation.

\begin{table}[!h]
\caption{Decay width for all three ensembles.}
\label{tab:width}
\vspace{2mm}
\begin{tabular}{cc}
\hline
Ensemble & $\Gamma_{\eta_c\gamma\gamma}/m_{\eta_c}\times 10^6$ \\
\hline
a98-I & 1.719(20) \\
a85-I & 1.847(19) \\
a67-I & 1.982(24) \\
a98-II & 1.701(20) \\
a85-II & 1.844(14) \\
a67-II & 1.986(15) \\
\hline
\end{tabular}

\end{table}

\begin{figure}[!h]
\centering
\subfigure{\includegraphics[width=0.47\textwidth]{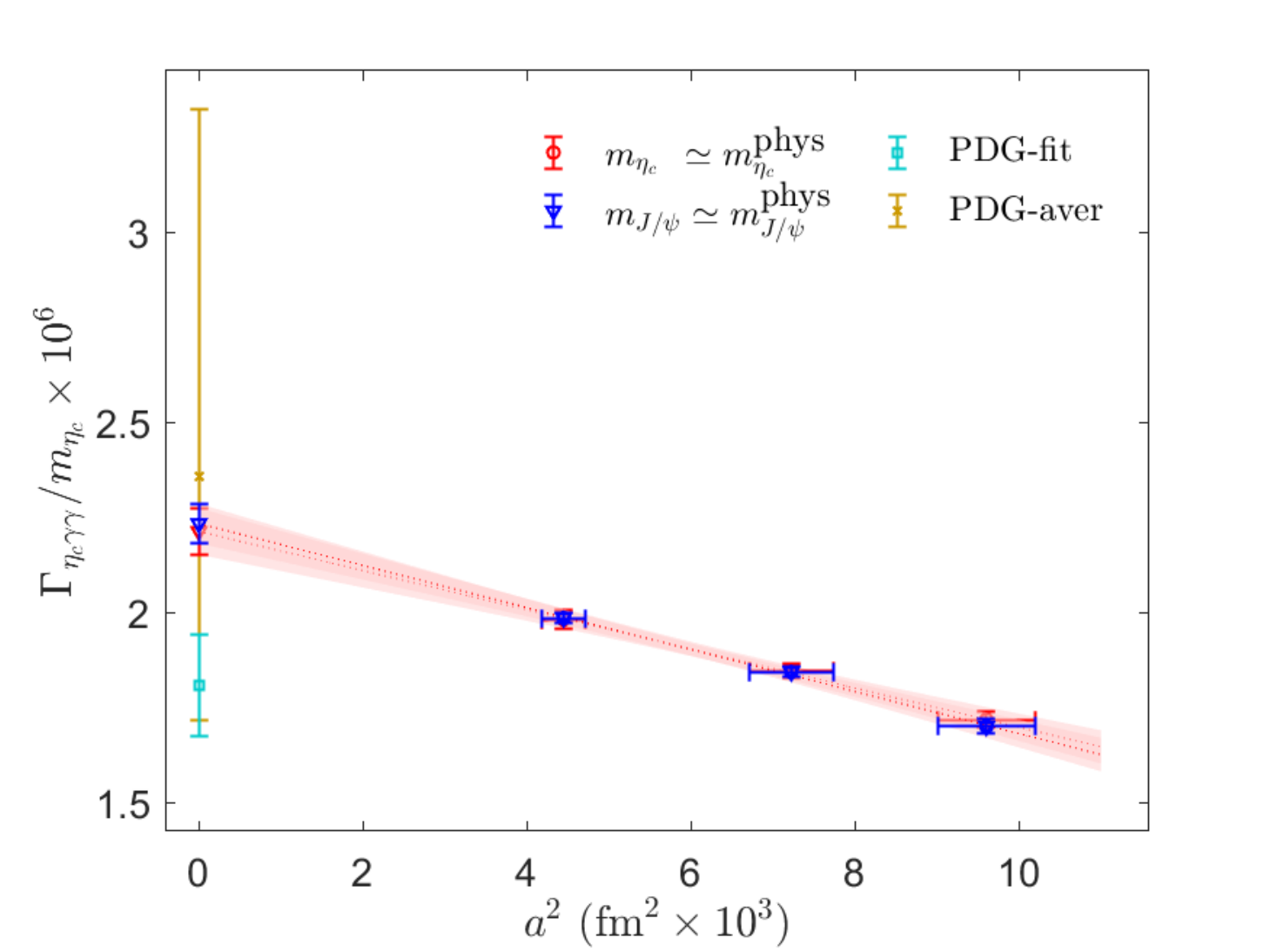}}\hspace{5mm}
\caption{(Color online) The lattice results of $\Gamma_{\eta_c\gamma\gamma}/m_{\eta_c}$ as a function of the lattice spacing. The errors of lattice spacing are presented by the horizontal error bars.
The symbol of the red circle and blue triangle denote the lattice results from ensemble-I and -II, respectively.
The symbols of the cyan square and orange cross indicate the PDG-fit and PDG-aver values.
Here PDG-aver data point is shifted
horizontally to favor a clear comparison.}
\label{fig:width_cont}
\end{figure}

Using the form factors as inputs, the dimensionless quantity $\Gamma_{\eta_c\gamma\gamma}/m_{\eta_c}$ can be evaluated and the results are listed in Table~\ref{tab:width}.
The lattice results for $\Gamma_{\eta_c\gamma\gamma}/m_{\eta_c}$ at different lattice spacings
are shown in Fig.~\ref{fig:width_cont} together with an extrapolation that is linear in $a^2$.
This is an expected behavior for the twisted mass ensembles which have the
automatic $O(a)$ improvement at maximal twist. It is evident that the fitting curves describe the lattice data well.
However, there is a possibility that ensemble a98 may not be optimally tuned, leading to residual $\mathcal{O}(a)$ discretization errors~\cite{ETM:2009ptp}.
To further explore this possibility, we perform the continuum
extrapolations both with and without the coarsest lattice, a98.
These results are consistent within statistical errors. After the continuous extrapolations of
dimensionless $\Gamma_{\eta_c\gamma\gamma}/m_{\eta_c}$, we obtain
\be
\label{eq:decay_width_ratio}
\frac{\Gamma_{\eta_c\gamma\gamma}}{m_{\eta_c}}=
\left\{
\begin{array}{lr}
2.214(61)\times 10^{-6} \; \quad \textrm{I, \;with a98}& \\
2.234(52)\times 10^{-6} \; \quad \textrm{II, with a98}&\\
2.211(100)\times 10^{-6} \quad \textrm{I, \;without a98}& \\
2.227(85)\times 10^{-6} \; \quad \textrm{II, without a98}&\\
\end{array}
\right.
\ee

To have a direct comparison with the experimental results of the decay width, we rescale the dimensionless $\Gamma_{\eta_c\gamma\gamma}/m_{\eta_c}$ to physical values by multiplying the experimental mass
of $\eta_c$, $m_{\eta_c}^{\textrm{exp}}$=2.9839 GeV. The following results for $\Gamma_{\eta_c\gamma\gamma}$ are then obtained:
\be
\label{eq:decay_width}
\Gamma_{\eta_c\gamma\gamma}=
\left\{
\begin{array}{lr}
6.61(18) \; \textrm{keV} \quad \textrm{I, \;with a98}& \\
6.67(16) \; \textrm{keV} \quad \textrm{II, with a98}&\\
6.60(30) \; \textrm{keV} \quad \textrm{I, \;without a98}& \\
6.64(25) \; \textrm{keV} \quad \textrm{II, without a98}&\\
\end{array}
\right.
\ee
The results of $\Gamma_{\eta_c\gamma\gamma}$ show good consistency across variations in the charm quark mass and with or without ensemble a98.
The latter suggests that no indication of $\mathcal{O}(a)$ effects are observed in ensemble a98, which is consistent with findings
from other lattice studies~\cite{ETM:2009ptp,Becirevic:2012dc,Alexandrou:2009qu,ETM:2009ztk}.
Therefore, we have included the results from ensemble a98 in our report.
Specifically, we use the central value and statistical error from the second line in Eq.~(\ref{eq:decay_width}) - which uses $J/\psi$ for the charm quark mass setting -
as the central value and statistical error for our final result. To estimate the systematic error, we consider the deviation of the central values between the first two lines in Eq.~(\ref{eq:decay_width}).
Our final result for the decay width is
\be
\label{eq:Gamma-final}
\Gamma_{\eta_c\gamma\gamma}=6.67(16)_{\mathrm{stat}}(6)_{\mathrm{syst}} \, \textrm{keV}.
\ee
This result is larger than the PDG-fit value by 24\% with a 2.9 $\sigma$ tension,
but compatible with PDG-aver value as it carries a much larger uncertainty.
Our lattice results using two different valence charm quark mass setting procedures,
together with the two PDG values are illustrated in Fig.~\ref{fig:width_cont}
for comparison.

It is worth noting that during the preparation of this paper, the authors of Ref.~\cite{Wang:2021dxw} reanalyze the
the experimental measurements related to $\eta_c$ decay after 1995, using the PDG-fit method. They reported the updated value as
$\Gamma_{\eta_c\gamma\gamma}=5.43
\begin{pmatrix}
\begin{smallmatrix}
+41 \\
-38
\end{smallmatrix}
\end{pmatrix}$ keV. The HPQCD collaboration~\cite{Colquhoun:2023zbc} also presents a lattice calculation using the traditional method and gives a result of $\Gamma_{\eta_c\gamma\gamma}=6.788(45)_{\textrm{fit}}(41)_{\textrm{syst}}$ keV, which agrees well with our calculation.

\section{Discussion}
The PDG-fit relies on the assumption that all individual measurements are correctly statistically distributed and the correlations among the different decay modes
are explicitly known. This is relatively difficult and the assumption
may not be valid for all the measurements.
According to the table of the correlation coefficients used in the PDG fit,
there exists a large correlation between the $\eta_c\rw2\gamma$ decay and other decay modes,
indicating that the PDG-fit result is easily affected by the parametrization in the constrained fit.
Thus a direct and precise experimental measurement of
$\eta_c\to2\gamma$ is essential for our better understanding of the charmonium radiative decays.
\begin{figure}[h]
\centering
\subfigure{\includegraphics[width=0.47\textwidth]{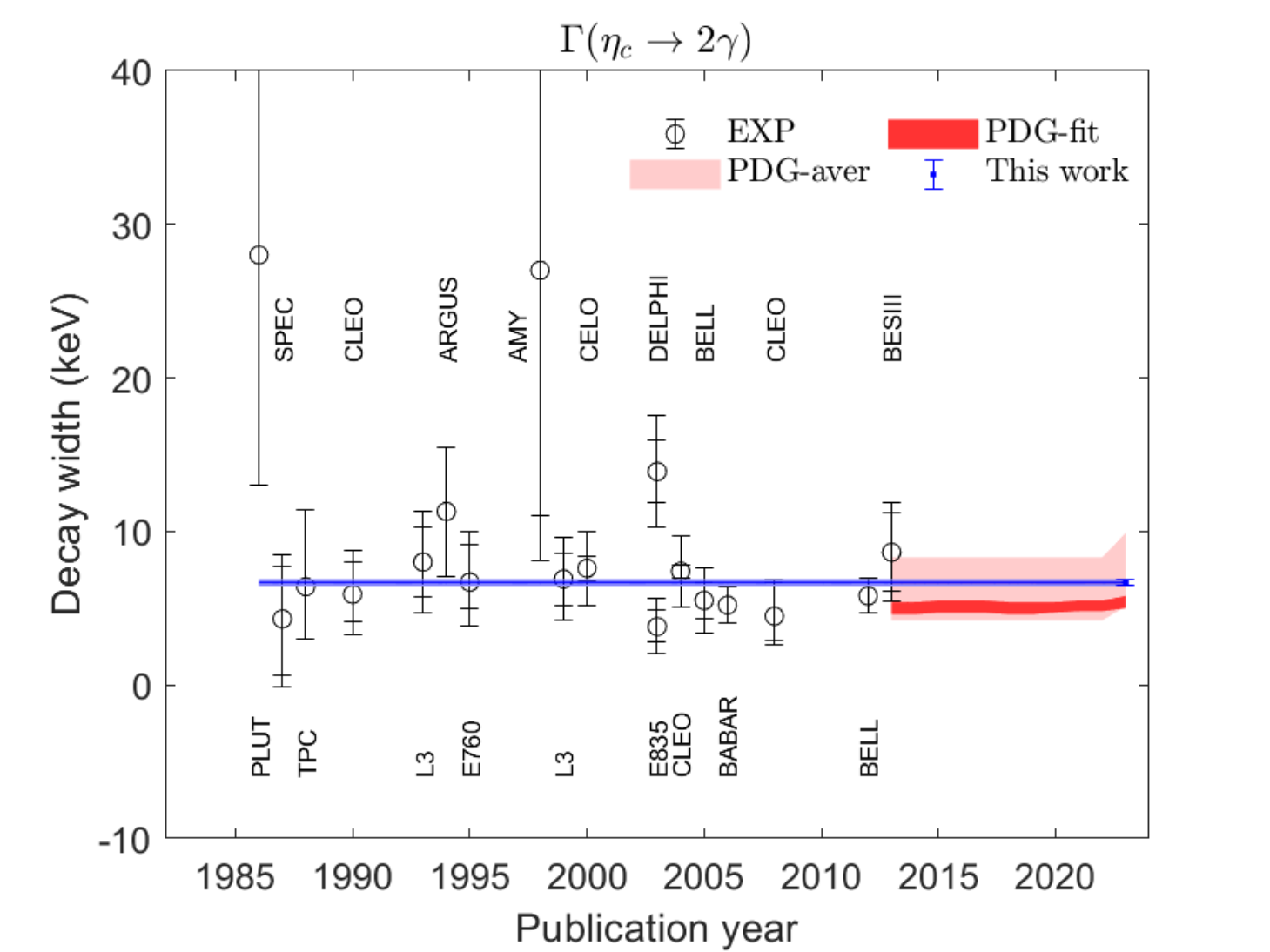}}\hspace{5mm}
\caption{(Color online) The historical evolution of the decay width $\Gamma(\eta_c\rw2\gamma)$ from
various experiments. Since 2013, PDG has started to produce the average and fit results using
BESIII (2013) and CLEO (2008) data as inputs. The lattice result, PDG-aver, and PDG-fit
results are represented by the blue band, light pink band, and deep red band, respectively.}
\label{fig:width_exp}
\end{figure}

In Fig.~\ref{fig:width_exp} we summarize the experimental measurements of $\Gamma_{\eta_c\gamma\gamma}$
from 1986 to 2013~\cite{PLUTO:1985fgr,Turin:1987qkd,TPCTwoGamma:1987tnp,CLEO:1989cel,L3:1993ahi,ARGUS:1994glp,E760:1995rep,AMY:1998ghf,L3:1999byj,CLEO:2000qcl,FermilabE835:2003ula,DELPHI:2003kmy,CLEO:2003gwz,Belle:2005fji,BaBar:2005pcw,Belle:2012uhr,CLEO:2008qfy,BESIII:2012lxx}. These results are very consistent with our lattice calculation but carry quite large errors, which range from 20\% to 100\%. It is still challenging to reduce the uncertainty to the level of a few percent.
Regarding this situation, for a certain period, a first-principle determination of $\Gamma_{\eta_c\gamma\gamma}$ from lattice QCD will play an irreplaceable role
for a better understanding of the QCD dynamics in charmonium physics.

In this work, we present a lattice calculation of $\eta_c\rightarrow 2\gamma$ with various systematic effects under well control using a novel method. The systematic error budget in our current lattice calculation is relatively complete. The remaining systematic effects requiring further investigation are the neglected disconnected diagrams, the quenching of strange quark, and the use of up and down quarks heavier than their physical values in our calculation. The first effect is Okubo-Zweig-Iizuka (OZI) suppressed and believed to only give a small contribution in the charmonium system~\cite{McNeile:2004wu,deForcrand:2004ia,Levkova:2010ft,Hatton:2020qhk}.
For the other two, previous lattice calculations~\cite{Bali:2011dc} indicate that they will also result in only small effects.
Nevertheless, These further improvements become more straightforward for future lattice calculation using the gauge ensembles with physical pion mass, heavy sea quarks, and more different lattice spacings. The calculation including the disconnected diagrams will also provide a direct estimation of the OZI suppressed contribution.

\section{Conclusion}
In this work, we propose a new method to compute the decay width
of $\eta_c\rw 2\gamma$, where the on-shell form factor
is obtained by combining the hadronic function calculated from lattice QCD
and an appropriate weight function known analytically.
As this method requires no model-dependent extrapolation of the off-shell form factors,
it provides a theoretically clean framework to determine the radiative decay width.
Such a method can also be applied to other processes which involve the leptonic or
radiative particles in the final states, for example $\pi^0 \rw 2\gamma$~\cite{Feng:2012ck,Gerardin:2016cqj},
$J/\psi \rw 3\gamma$~\cite{Meng:2019lkt}, $\pi\to e^+e^-$~\cite{Christ:2020dae,Christ:2022rho}, $K_L\to\mu^+\mu^-$~\cite{Christ:2020bzb}
and radiative leptonic decays $K^-\to\ell^-\bar{\nu}\gamma$, $D_s^+\to \ell^+\nu\gamma$,
$B\to\ell^-\bar{\nu}\gamma$~\cite{Kane:2019jtj,Desiderio:2020oej}.
We have also paid special attention to other systematic effects such as finite volume effects,
excited states contaminations, tuning of the valence charm quark mass
and finally, quite substantial finite lattice spacing errors have been
observed in previous lattice calculations. Taking into account the improvements mentioned above,
we managed to obtain a result for the decay rate in Eq.~(\ref{eq:Gamma-final}) with an error of about 2.6\%.

Our lattice result for the decay width is consistent with most of the previous experimental measurements
and also the PDG-aver value, but notably different from
the PDG-fit value by about 2.9 standard deviations.
We suspect that the error of the PDG-fit value might be greatly underestimated
due to the limited knowledge of the direct experimental
measurements of $\eta_c \rw 2\gamma$ and strong correlation from other decay channels.
It is therefore crucial for the forthcoming experiments, e.g. BESIII,
to further reduce the experimental uncertainties.
The cross-check of our results by other lattice QCD calculations are also very helpful
to clarify the discrepancy between theory and experiments.

\section{Acknowledgments}
We thank ETM Collaboration for sharing the gauge configurations with us. A particular acknowledgment goes to Carsten Urbach.
We gratefully acknowledge many helpful discussions with Michael Doser, Luchang Jin, Haibo Li, and Yan-Qing Ma. We thank Christine Davies for reminding us the PDG values are updated.
Y.M. acknowledges support by the National Natural Science Foundation
of China (12047505) and
State Key Laboratory of Nuclear Physics and Technology, Peking University.
X.F. and T.W. were supported in part by the National Natural Science Foundation
of China (12125501, 12141501, and 11775002), and National Key Research and Development Program of China (2020YFA0406400).
X.F., C.L., and Z.H.Z. are supported in part by the National Natural Science Foundation
of China (12070131001).
C.L. and Z.H.Z. are also supported in part by CAS Interdisciplinary Innovation Team and the National Natural Science Foundation
of China (11935017, 12293060, and 12293063).
The main calculation was carried out on Tianhe-1A supercomputer at Tianjin National
Supercomputing Center and partly supported by High-performance Computing Platform
of Peking University.

% \bibliography{etac_2photon}
%\bibliography{ref}
% \bibliographystyle{unsrt
%\bibliographystyle{model3-num-names}

\clearpage

\section{Supplementary Material}
In this section, we expand on a selection of technical details and add
results to facilitate cross-checks of different calculations of
the $\eta_c\to2\gamma$ decay width.

\subsection{Derivation of Eq.~(5)}
For convenience, we introduce a scalar function $\mathcal{I}$ by multiplying $\epsilon_{\mu\nu\alpha \beta}p_{\alpha}k_\beta$ to the both sides of the Eq.~(5). For the right side, it has
\beq
\mathcal{I}&\equiv& \epsilon_{\mu\nu\alpha \beta}p_{\alpha}k_\beta \epsilon_{\mu\nu\alpha' \beta'}p_{\alpha'}k_{\beta'}F_{H\gamma\gamma} \nonumber \\
&=& 2F_{H\gamma\gamma}[p^2k^2-(p\cdot k)^2] \nonumber \\
&=&-2F_{H\gamma\gamma}m_{H}^2|\vec{p}|^2 
\eeq
where $|\vec{p}|=m_{H}/2$ is considered. For the left side, it has

\beq\label{eq:S_2}
\mathcal{I}
&=&im_H\epsilon_{\mu\nu\alpha 0}p_{\alpha}\int dt\,e^{m_Ht/2}\int d^3 \vec{x}\,
e^{-i\vec{p}\cdot \vec{x}}\mathcal{H}_{\mu\nu}(t,\vec{x}) \nonumber \\
&=&im_H\epsilon_{\mu\nu\alpha 0}\int dt\,e^{m_Ht/2}\int d^3 \vec{x}\,
\left(i\frac{\partial}{\partial x_{\alpha}}e^{-i\vec{p}\cdot \vec{x}}\right) \mathcal{H}_{\mu\nu}(t,\vec{x}) \nonumber \\
&=&-im_H\epsilon_{\mu\nu\alpha 0}\int dt\,e^{m_Ht/2}\int d^3 \vec{x}
e^{-i\vec{p}\cdot \vec{x}} i\frac{\partial}{\partial x_{\alpha}}\mathcal{H}_{\mu\nu}(t,\vec{x}) \nonumber \\
&=&m_H\int dt\,e^{m_Ht/2}\int d^3 \vec{x}
e^{-i\vec{p}\cdot \vec{x}} \epsilon_{\mu\nu\alpha 0} \frac{\partial}{\partial x_{\alpha}}\mathcal{H}_{\mu\nu}(t,\vec{x})
\eeq
Combining these two equations, we obtain
\be
F_{H\gamma\gamma}=-\frac{1}{2m_{H}|\vec{\bm{p}}|^2}\int d^4x\,e^{-ipx}
\epsilon_{\mu\nu\alpha0}
\frac{\partial \mathcal{H}_{\mu\nu}(x) }{\partial x_{\alpha}}.
\ee

Averaging the direction of $\vec{p}$, then the factor $e^{-i\vec{p}\cdot \vec{x}}$ can be replaced by
\beq
\frac{1}{4\pi}\int d\Omega_{\vec{p}}e^{-i\vec{\bm{p}}\cdot \vec{\bm{x}}}&=& \frac{1}{2}\int_{-1}^{1} d\cos\theta e^{-i|\vec{\bm{p}}||\vec{\bm{x}}|\cos\theta} \nonumber \\ 
&=& \frac{\sin(|\vec{p}||\vec{x}|)}{|\vec{p}||\vec{x}|}\equiv j_0(|\vec{p}||\vec{x}|)
\eeq
Therefore, 
\beq\label{eq:S_5}
F_{H\gamma\gamma}&=&-\frac{1}{2m_{H}|\vec{\bm{p}}|^2}\int dt\,e^{m_Ht/2}\int d^3 \vec{x} \nonumber \\
&\times& j_0(|\vec{p}||\vec{x}|)\epsilon_{\mu\nu\alpha0}
\frac{\partial \mathcal{H}_{\mu\nu}(x) }{\partial x_{\alpha}} \, \nonumber \\
&=&\frac{1}{2m_{H}|\vec{\bm{p}}|^2}\int dt\,e^{m_Ht/2}\int d^3 \vec{x} \nonumber \\
&\times& \epsilon_{\mu\nu\alpha0}
\frac{\partial j_0(|\vec{p}||\vec{x}|) }{\partial x_{\alpha}}\mathcal{H}_{\mu\nu}(x) \nonumber \\
&=&-\frac{1}{2m_{H}}\int d^4x\, e^{\frac{m_{H}}{2}t}\,
\frac{j_1(|\vec{\bm{p}}||\vec{\bm{x}}|)}{|\vec{\bm{p}}||\vec{\bm{x}}|}\,
\epsilon_{\mu\nu\alpha0}x_{\alpha}\mathcal{H}_{\mu\nu}(x), \nonumber \\
\eeq
Both in the penultimate lines of Eq.~\ref{eq:S_2} and Eq.~\ref{eq:S_5}, the space integrals for the total differentiation are omitted.

\subsection{Hadroinc function $\mathcal{H}_{\mu\nu}(t,\vec{x})$}

The hadroinc function $\mathcal{H}_{\mu\nu}(t,\vec{x})$ in Eq.~(2)  can be
extracted from a three-point function $C_{\mu\nu}(t,\vec{x};\Delta t)$
\be
C_{\mu\nu}(t,\vec{x};\Delta t)=
\left\{
             \begin{array}{lr}
             \langle J_{\mu}^{em}(\vec{x},t)J_{\nu}^{em}(\vec{0},0) \phi_{\eta_c}^{\dagger}(-\Delta t) \rangle, &t \geq 0 \\
             \langle J_{\mu}^{em}(\vec{0},0)J_{\nu}^{em}(\vec{x},t) \phi_{\eta_c}^{\dagger}(t-\Delta t) \rangle, &t < 0 \\

             \end{array}
\right.
\ee
with only the connected Wick contractions included
\beq
 &&\langle J_{\mu}^{em}(\vec{x},t)J_{\nu}^{em}(\vec{0},0) \phi_{\eta_c}^{\dagger}(-\Delta t) \rangle \nonumber \\
 &&=-Z_V^2e_c^2\langle\bar{c}\gamma_{\mu}c(\vec{x},t)\bar{c}\gamma_{\nu}c(\vec{0},0)\bar{c}\gamma_5c(-\Delta t) \rangle \nonumber \\
 &&=Z_V^2e_c^2\Big{\{}\langle \textrm{Tr}[\gamma_5S_c(-\Delta t;\vec{x},t)\gamma_{\mu}S_c(\vec{x},t;\vec{0},0)
 \nonumber \\
 &&\times \gamma_{\nu}S_c(\vec{0},0; -\Delta t) ]\rangle +\langle \textrm{Tr}[\gamma_5S_c(-\Delta t;\vec{0},0)\nonumber \\
 &&\times \gamma_{\nu}S_c(\vec{0},0;\vec{x},t)
 \gamma_{\mu}S_c(\vec{x},t; -\Delta t) ]\rangle \Big{\}}
\eeq
and
\beq
 &&\langle J_{\mu}^{em}(\vec{0},0)J_{\nu}^{em}(\vec{x},t) \phi_{\eta_c}^{\dagger}(t-\Delta t) \rangle \nonumber \\
 &&=-Z_V^2e_c^2\langle\bar{c}\gamma_{\mu}c(\vec{0},0)\bar{c}\gamma_{\nu}c(\vec{x},t)\bar{c}\gamma_5c(t-\Delta t) \rangle \nonumber \\
 &&=Z_V^2e_c^2\Big{\{}\langle \textrm{Tr}[\gamma_5S_c(t-\Delta t;\vec{0},0)\gamma_{\mu}S_c(\vec{0},0;\vec{x},t) \nonumber \\
 &&\times \gamma_{\nu}S_c(\vec{x},t; t-\Delta t) ]\rangle +\langle \textrm{Tr}[\gamma_5S_c(t-\Delta t;\vec{x},t)\nonumber \\
&& \times \gamma_{\nu}S_c(\vec{x},t;\vec{0},0)
 \gamma_{\mu}S_c(\vec{0},0; t-\Delta t) ]\rangle \Big{\}}
\eeq

Then, the hadroinc function $\mathcal{H}_{\mu\nu}(t,\vec{x})$ is determined directly through
\be
\mathcal{H}_{\mu\nu}(t,\vec{x})=
\left\{
             \begin{array}{lr}
             \frac{2E_0}{Z_0}e^{E_0\Delta t}C_{\mu\nu}(t,\vec{x};\Delta t), &t \geq 0 \\
             \frac{2E_0}{Z_0}e^{E_0(\Delta t-t)}C_{\mu\nu}(t,\vec{x};\Delta t), &t < 0 \\

             \end{array}
\right.
\ee
where $E_0, Z_0$ are extracted from the two point function as given in Eq.~(9) and
the renormalization constant $Z_V$ is calculated by Eq.~(\ref{eq:zv_ratio}).

\subsection{Tuning of the valence charm quark mass}

The detailed information of the charm quark mass tuning is given in Table~\ref{tab:charm_quark_mass}.
We attach the suffix ``-I'' and ``-II'' to distinguish the cases with $m_{\eta_c}\simeq m_{\eta_c}^{\mathrm{phys}}$ and $m_{J/\psi}\simeq m_{J/\psi}^{\mathrm{phys}}$.
Together with the charmonium masses, we also list the values of the hyperfine splitting $\delta m\equiv m_{J/\psi}-m_{\eta_c}$. Using three ensembles, an extrapolation that is linear in $a^2$ can be performed, see Fig.~\ref{fig:hyperfine_splitting}), and in the continuum limit we obtain $\delta m=123(1)$ MeV,
which is 10 MeV larger than the PDG value. Similar increase of the hyperfine splitting
has been observed by the HPQCD collaboration~\cite{Hatton:2020qhk}, 
 where the discarded $\eta_c$ annihilation effects
are expected to cause a $7.3(1.2)$ MeV enhancement in $\delta m$.
In our calculation, a similar shift in $\delta m$ could lead to a $\sim0.3\%$ 
 uncertainty in the $\eta_c$ mass, which has little effect on our final result for
 the decay rate in Eq.~(12).

\begin{table}[!h]
\begin{ruledtabular}
\begin{tabular}{cccccc}
& \textrm{Ensemble} & $a\mu_{c}$ & $m_{\eta_c}(\textrm{MeV})$  & $m_{J/\psi}(\textrm{MeV})$
& $\frac{\delta m}{m_{\eta_c}}\times 10^{2}$  \\
%\multicolumn{1}{c}{\textrm{Three}}& \\
\hline
& a98-I             &   0.2896  & 2984.9(4) & 3064.8(5) &2.677(12)      \\
& a85-I            &   0.2563  & 2988.6(5) & 3078.0(7)   &2.991(18)  \\
& a67-I            &    0.2024   & 2987.6(5) & 3090.4(7) &3.434(15)  \\
& Cont.Limt & --- & --- & --- & 4.11(11)\\
\hline
& a98-II           &  0.2951   & 3018.9(4) & 3097.2(5)   &2.596(11)   \\
& a85-II          &  0.2586   & 3005.7(5) & 3094.4(7)    &2.953(18)   \\
& a67-II          &   0.2036   & 2999.5(5) & 3101.7(7)   &3.411(15) \\
& Cont.Limt & --- & --- & --- & 4.13(12)\\
\hline
& PDG & --- & 2983.9(4) & 3096.9(0) & 3.79(1)\\
\end{tabular}
\end{ruledtabular}
\caption{
Tuning of the bare charm quark mass $\mu_c$. The uncertainties of $m_{\eta_c}$ and $m_{J/\psi}$ are statistical only. The mass $\mu_c$ is tuned such that the mass of $\eta_c$ or $J/\psi$ approaches to its physical value, with the deviation controlled to be less than 0.2\%. To distinguish the cases with $m_{\eta_c}\simeq m_{\eta_c}^{\mathrm{phys}}$ and $m_{J/\psi}\simeq m_{J/\psi}^{\mathrm{phys}}$,
we attach the suffix ``-I'' and ``-II'' to the ensemble name. $\delta m\equiv m_{J/\psi}-m_{\eta_c}$ 
designates the hyperfine splitting of the charmonia.}
\label{tab:charm_quark_mass}
\end{table}

\renewcommand\thefigure{\Alph{section}S\arabic{figure}}    
\setcounter{figure}{0}   
\begin{figure}[h]
	\centering
		\subfigure{\includegraphics[width=0.48\textwidth]{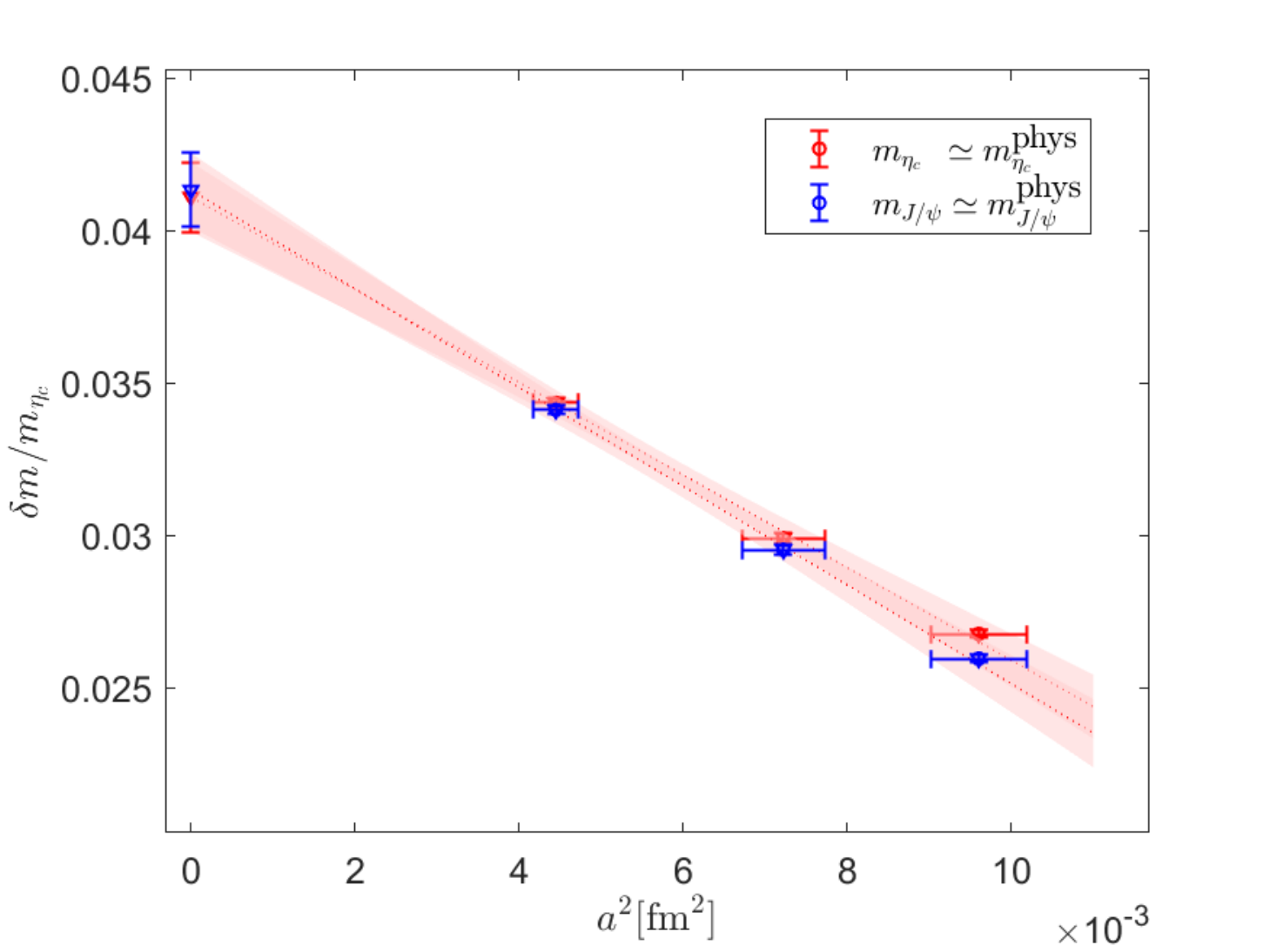}}\hspace{5mm}
    \caption{(Color online) The continuum extrapolation of hyperfine splitting. The errors of lattice spacing are presented by the  horizontal error bars.}
    \label{fig:hyperfine_splitting}
\end{figure}

\subsection{Determination of $Z_V$}

In our calculation the electromagnetic current is replaced by a local charm quark current as
$J^{em}_\mu(x)=Z_V e_cJ^{(c)}_{\mu}(x)$ with $J^{(c)}_{\mu}$ defined as
$J^{(c)}_{\mu}\equiv \bar{c}\gamma_{\mu}c$, at the cost of introducing at most $O(a^2)$ errors,
most of which are taken care of by the continuum extrapolation procedure described in main body of this paper. The factor $Z_V$ is a vector-current renormalization factor, which can be
calculated by applying the condition of charge conservation and using a ratio between $C^{(2)}(t)$
and the three-point function $C^{(3)}(t)=\sum_{\vec{x}}\langle \mathcal{O}_{\eta_c}(t) J^{(c)}_{0}(t/2,\vec{x}) \mathcal{O}_{\eta_c}^\dagger(0)\rangle$
with zero three-momentum inserted for both initial and final states. As the charge conservation holds for both ground and excited-states,
we find that the excited-state effects in $C^{(3)}(t)$ and $C^{(2)}(t)$ cancel efficiently. The plateau of the ratio starts at $t\approx 1$ fm.
The main systematic effect appears as
the around-of-world effect in $C^{(2)}(t)$ at $t\approx T/2$. To account for this effect, we use the following ansatz, 
\be
\label{eq:zv_ratio}
Z_V=\frac{C^{(2)}(t)}{C^{(3)}(t)}\frac{1}{\left(1+e^{-m_{\eta_c}(T-2t)}\right)},\quad \mbox{for }t\lesssim T/2,
\ee
to extract $Z_V$ from the ratio.
We find that the uncertainty in the determination of the charm quark mass makes a nearly negligible impact on $Z_V$, whose numerical values are listed in Table~\ref{tab:ZV}. We have also checked that 
the values for $Z_V$ presented here are consistent with those calculated by the RI-MOM scheme in Ref.~\cite{ETM:2010iwh}, where the results are given as 0.604(07), 0.624(04) and 0.659(04) for $a=0.098,0.085,0.0667$ fm, respectively.

\begin{table}[!h]
\begin{ruledtabular}
\begin{tabular}{cccc}
 Ensemble & a98-I & a85-I & a67-I   \\
\hline
$Z_V$   &   0.6033(20)  & 0.6255(22) & 0.6517(15)   \\
\hline
Ensemble & a98-II  &  a85-II & a67-II  \\
\hline
$Z_V$  & 0.6047(19)  & 0.6257(21) & 0.6516(15)   \\
\end{tabular}
\end{ruledtabular}
\caption{The vector-current renormalization constants $Z_V$ calculated using Eq.~(\ref{eq:zv_ratio}).}
\label{tab:ZV}
\end{table}

\bibliography{ref}

\begin{thebibliography}{99}


\bibitem{Brambilla:2010cs}
Brambilla N, Eidelman S, Heltsley B-K, et al. Heavy quarkonium: progress, puzzles, and opportunities. Eur Phys J C 2011; 71:1534.

\bibitem{Bodwin:1994jh}
Geoffrey T-B, Eric B, Peter L-B. Rigorous QCD analysis of inclusive annihilation and production of heavy quarkonium. Phys Rev D 1995;51:1125 [Erratum: Phys Rev D 1997;55:5853 (1997)].

\bibitem{Eichten:1995ch}
Eichten E-J, Quigg C. Quarkonium wave functions at the origin. Phys Rev D 1995;52:1726.

\bibitem{PLUTO:1985fgr}
PLUTO Collaboration, Berger C, et al. Evidence for exclusive $\eta_c$ production in $\gamma \gamma$ interactions. Phys Lett B 1986;167:120.

\bibitem{Turin:1987qkd}
ACGLORST Collaboration, Baglin C, et al. Direct observation and partial width measurement of $\gamma \gamma$ decay of charmonium states. Phys Lett B 1987;187:191.

\bibitem{TPCTwoGamma:1987tnp}
TPC/Two-Gamma Collaboration, Aihara H, et al. Charmonium production in photon-photon collisions. Phys Rev Lett 1988;60:2355.

\bibitem{CLEO:1989cel}
CLEO Collaboration, Chen W-Y, et al. Measurement of $\gamma \gamma$ widths of charmonium states. Phys Lett B 1990;243:169.

\bibitem{L3:1993ahi}
L3 Collaboration, Adriani O, et al. Measurement of $\eta_c$ production in untagged two photon collisions
at LEP. Phys Lett B 1993;318:575.

\bibitem{ARGUS:1994glp}
ARGUS Collaboration, Albrecht H, et al. Determination of the radiative decay width of the $\eta_c$ meson. Phys Lett B 1994;338:390.

\bibitem{E760:1995rep}
E760 Collaboration, Armstrong T-A, et al. Study of the $\eta_c(1^1S_0)$ state of charmonium formed in
$\bar{p}p$ annihilations and a search for the $\eta_c^{\prime}(2^1S_0)$.
Phys Rev D 1995;52:4839.

\bibitem{AMY:1998ghf}
AMY Collaboration, Shirai M, et al. Observation of exclusive $\eta_c$ production in two photon interactions at TRISTAN. Phys Lett B 1998;424:405.

\bibitem{L3:1999byj}
L3 Collaboration, Acciarri M, et al. Formation of the $\eta_c$ in two photon collisions at LEP. Phys Lett B 1999;461:155.

\bibitem{CLEO:2000qcl}
CLE0 Collaboration, Brandenburg G, et al. Measurements of the mass, total width and two photon partial width
of the $\eta_c$ meson. Phys Rev Lett 2000;85:3095.

\bibitem{FermilabE835:2003ula}
Fermilab E835 Collaboration, Ambrogiani M, et al. Measurement of the resonance parameters of the charmonium ground
state, $\eta_c(1^1S_0)$. Phys Lett B 2003;566:45.

\bibitem{DELPHI:2003kmy}
DELPHI Collaboration, Abdallah J, et al. The $\eta_c(2980)$ formation in two photon collisions at LEP energies. Eur Phys J C 2003;31:481.

\bibitem{CLEO:2003gwz}
CLEO Collaboration, Asner D-M, et~al. Observation of $\eta_c'$ production in $\gamma\gamma$ fusion at
CLEO. Phys Rev Lett 2004;92:142001.

\bibitem{Belle:2005fji}
BELLE Collaboration, Kuo C-C, et al. Measurement of $\gamma\gamma\rw \bar{p}p$ production at Belle. Phys Lett B 2005;621:41.

\bibitem{BaBar:2005pcw}
BARBAR Collaboration, Aubert B, et al. Measurements of the absolute branching fractions of $B^{\pm}\rw K^{\pm}X_{c\bar{c}}$. Phys Rev Lett 2006;96:052002.

\bibitem{Belle:2012uhr}
BELLE Collaboration, Zhang C-C, et al. First study of $\eta_c$, $\eta(1760)$ and $X(1835)$ production via $\eta^\prime\pi^+\pi^-$ final states in two-photon collisions. Phys Rev D 2012;86:052002.

\bibitem{BESIII:2012lxx}
BESIII Collaboration, Ablikim M, et al. Evidence for $\eta_c\rw 2\gamma$ and measurement of $J/\psi \rw 3\gamma$. Phys Rev D 2013;87:032003.

\bibitem{CLEO:2008qfy}
CLEO Collaboration, Adams G-S, et al. Observation of $J/\psi\rw 3\gamma$.
Phys Rev Lett 2008;101:101801.

\bibitem{Godfrey:1985xj}
Godfrey S, Isgur N, Mesons in a relativized quark model with chromodynamics. Phys Rev D 1985;32:189.

\bibitem{Dudek:2006ut}
Dudek J-J, Edwards R-G. Two photon decays of charmonia from lattice QCD. Phys Rev Lett 2006; 97:172001.

\bibitem{Feng:2015uha}
Feng F, Jia Y, Sang W-L. Can nonrelativistic QCD explain the $\gamma\gamma^* \rw \eta_c$
transition form Factor data? Phys Rev Lett 2015;115:222001.

\bibitem{CLQCD:2016ugl}
CLQCD Collaboration, Chen T, et al. Two-photon decays of $\eta _c$ from lattice QCD. Eur Phys J C 2016;76:358.

\bibitem{Chen:2016bpj}
Chen J, Ding M-H, Chang L, et al. Two photon transition form factor of $\bar{c}c $ Quarkonia.
Phys Rev D 2017; 95:016010.

\bibitem{Feng:2017hlu}
Feng F, Jia Y, Sang W-L. Next-to-next-to-leading-order QCD corrections to the hadronic width
of pseudoscalar quarkonium. Phys Rev Lett 2017;119:252001.

\bibitem{Li:2019ncs}
Li R, Feng Y, Ma Y-Q. Exclusive quarkonium production or decay in soft gluon
factorization. J High Energy Phys 2020;05:009.

\bibitem{Liu:2020qfz}
Liu C, Meng Y, Zhang K-L. Ward identity of the vector current and the decay rate of
$\eta_c\rightarrow\gamma\gamma$ in lattice QCD. Phys Rev D 2020;102:034502.

\bibitem{Zhang:2021xvl}
Zhang R-Q, Sun W, Chen Y, et al. The glueball content of $\eta_c$. Phys Lett B 2022; 827:136960.

\bibitem{Feng:2018qpx}
Feng X, Jin L-C. QED self energies from lattice QCD without power-law finite-volume errors. Phys Rev D 2019;100:094509.

\bibitem{Tuo:2019bue}
Tuo X-Y, Feng X, Jin L-C. Long-distance contributions to neutrinoless double beta decay $\pi^-
\rw \pi^+ e e$. Phys Rev D 2019;100:094511.

\bibitem{Feng:2019geu}
Feng X, Fu Y, Jin L-C. Lattice QCD calculation of the pion charge radius using a
model-independent method. Phys Rev D 2020;101:051502.

\bibitem{Feng:2020zdc}
Feng X, Gorchtein M, Jin L-C, et al. First-principles calculation of electroweak box diagrams from
lattice QCD. Phys Rev Lett 2020;124:192002.

\bibitem{Christ:2020jlp}
Christ N-H, Feng X, Jin L-C, et al. Electromagnetic corrections to leptonic pion decay from lattice QCD using infinite-volume reconstruction method. PoS LATTICE 2020;2019:259.

\bibitem{Christ:2020hwe}
Christ N-H, Feng X, Jin L-C, et al. Finite-volume effects in long-distance processes with massless
leptonic propagators. Phys Rev D 2021;103:014507.

\bibitem{Ma:2021azh}
Ma P-X, Feng X, Gorchtein M, et al. Lattice QCD calculation of the electroweak box diagrams for the kaon semileptonic decays. Phys Rev D 2021;103:114503.

\bibitem{Tuo:2021ewr}
Tuo X-Y, Feng X, Jin L-C, et al. Lattice QCD calculation of $K\to \ell\nu_\ell \ell'^+ \ell'^-$ decay width. Phys Rev D 2022;105:054518.

\bibitem{Feng:2021zek}
Feng X, Jin L-C, Riberdy M-J. Lattice QCD calculation of the pion mass splitting.
Phys Rev Lett 2022;128:052003.

\bibitem{Fu:2022fgh}
Fu Y, Feng X, Jin L-C, et al. Lattice QCD calculation of the two-photon exchange contribution to
the muonic-hydrogen lamb shift. Phys Rev Lett 2022;128:172002.

\bibitem{Tuo:2022hft}
Tuo X-Y, Feng X, Jin L-C. Lattice QCD calculation of the light sterile neutrino contribution
in 0\ensuremath{\nu}2\ensuremath{\beta} decay.
Phys Rev D 2022;106:074510.

\bibitem{Becirevic:2012dc}
Becirevic D, Sanfilippo F. Lattice QCD study of the radiative decays $J/\psi\rw \eta_c\gamma$
and $h_c\rw \eta_c\gamma$. J High Energy Phys 2013;01:028.

\bibitem{ETM:2009ptp}
ETM Collaboration, Blossier B, et al. Pseudoscalar decay constants of kaon and D-mesons from $N_f=2$ twisted mass lattice QCD. J High Energy Phys 2009;07:043.

\bibitem{APE:1987ehd}
APE Collaboration, Albanese M, et al. Glueball masses and string tension in lattice QCD.
Phys Lett B 1987;192:163.

\bibitem{Gusken:1989qx}
G{\"u}sken S. A study of smearing techniques for hadron correlation functions. Nucl Phys B Proc Suppl 1990;17:361.

\bibitem{Alexandrou:2009qu}
Alexandrou C, Baron R, Carbonell J, et al. Low-lying baryon spectrum with two dynamical twisted mass fermions. Phys Rev D 2009;80:114503.

\bibitem{ETM:2009ztk}
ETM Collaboration, Baron R, et al. Light meson physics from maximally twisted mass lattice QCD.
J High Energy Phys 2010;08:097.

\bibitem{Wang:2021dxw}
Wang H-P, Yuan C-Z. New puzzle in charmonium decays.
Chin Phys C 2022;46:071001.

\bibitem{Colquhoun:2023zbc}
Colquhoun B, Cooper L-J, Christine T-H, Precise determination of decay rates for $\eta_c \rw \gamma \gamma$, $J/\psi \rw \gamma \eta_c$ and $J/\psi \rw \eta_c e^+e^-$ from lattice QCD, arXiv:2305.06231,2023.

\bibitem{McNeile:2004wu}
McNeile C, Michael C. An estimate of the flavor singlet contributions to the hyperfine
splitting in charmonium. Phys Rev D 2004;70:034506.

\bibitem{deForcrand:2004ia}
Forcrand P-D, Perez M-G, Matsufuru H, et al. Contribution of disconnected diagrams to the hyperfine splitting of charmonium. J High Energy Phys 2004;08:004.

\bibitem{Levkova:2010ft}
Levkova L, DeTar C. Charm annihilation effects on the hyperfine splitting in charmonium. Phys Rev D 2011;83:074504.

\bibitem{Hatton:2020qhk}
Hatton D, Davies C-T-H, Galloway B, et al. Charmonium properties from lattice QCD+QED : hyperfine splitting, $J/\psi$ leptonic width, charm quark mass, and $a^c_\mu$. Phys Rev D 2020;102:054511.

\bibitem{Bali:2011dc}
Bali G, Collins S, Durr S, et al. Spectra of heavy-light and heavy-heavy mesons containing charm quarks, including higher spin states for $N_f=2+1$. PoS LATTICE 2021;2011:135.

\bibitem{Feng:2012ck}
Feng S, Aoki S, Fukaya H, et al. Two-photon decay of the neutral pion in lattice QCD.
Phys Rev Lett 2012;109:182001.

\bibitem{Gerardin:2016cqj}
G\'erardin A, Meyer H-B, Nyffeler A. Lattice calculation of the pion transition form factor $\pi^0 \rw \gamma^* \gamma^*$. Phys Rev D 2016;94:074507.

\bibitem{Meng:2019lkt}
Meng Y, Liu C, Zhang K-L. Three photon decay of $J/\psi$ from lattice QCD.
Phys Rev D 2020;102:054506.

\bibitem{Christ:2020dae}
Christ N-H, Feng X, Jin L-C, et al. Calculating the two-photon contribution to $\pi^0 \rightarrow e^+ e^-$ decay amplitude. PoS LATTICE 2020;2019:097.

\bibitem{Christ:2022rho}
Christ N-H, Feng X, Jin L-C, et al. Lattice QCD calculation of $\pi^0\rw e^+e^-$ decay. Phys Rev Lett 2023;130:191901.

\bibitem{Christ:2020bzb}
Christ N-H, Feng X, Jin L-C, et al. Lattice QCD calculation of the two-photon contributions to $K_L \rw \mu^+ \mu^-$ and $\pi^0 \rw e^+ e^-$ decays. PoS LATTICE 2020;2019:128.

\bibitem{Kane:2019jtj}
Kane C, Lehner C, Meinel S, et al. Radiative leptonic decays on the lattice. PoS LATTICE 2019;2019:134.

\bibitem{Desiderio:2020oej}
Desiderio A, Frezzotti R, Garofalo M, et al. First lattice calculation of radiative leptonic decay rates of
pseudoscalar mesons. Phys Rev D 2021;103:014502.

\end{thebibliography}

\bibliographystyle{h-physrev}

\end{document}